# A Robust Fiber-based Frequency Synchronization System Immune to Dramatic Temperature Fluctuation

Xi Zhu, Bo Wang, Yichen Guo, Yibo Yuan, Romeo Gamatham, Bruce Wallace, Keith Grainge, and Lijun Wang

*Abstract*—Fiber-based frequency synchronization system is sensitive to temperature change because of the limited isolation and nonlinear effect of RF components in the system. In order to make it suitable for the use of large-scale scientific and engineering projects in which the ambient temperature of the fiber link changes dramatically, we designed a non-harmonic frequency dissemination system immune to temperature fluctuation. After the lab tests in which the ambient temperature of fiber fluctuates 40°C/day and 20°C/hour respectively, the relative frequency stabilities of this system reached $4.0\times10^{-14}$/s and $3.0\times10^{-16}/10^4$s. It is demonstrated that the proposed non-harmonic scheme shows a strong robustness to complicated working environment with dramatic temperature fluctuation.

## I. Introduction

As the world's largest radio telescope under construction, the Square Kilometre Array (SKA) consists of thousands of parabolic dishes and aperture arrays which work together to form a one-square-kilometer collecting area. SKA will surpass any current telescopes and help us understanding the universe further. To ensure sufficient imaging fidelity, all of the telescopes should keep time and frequency synchronized with stabilities superior to $2.31\times10^{-12}$/s, $3.84\times10^{-14}$/min and $1.92\times10^{-14}$/10min [1]. In 2015, we proposed and demonstrated a precision reference frequency synchronization scheme via 1f-2f dissemination according to requirements and characteristics of SKA [2], we name it as harmonic system. Considering our previous work along with the existing fiber-based time and frequency transfer schemes demonstrated by different groups[3-14], they almost utilized fiber spools or urban telecommunication fibers which basically run in buried cables. While for large-scale scientific and engineering projects such as SKA, on account of attractive infrastructure cost and simple laying, overhead fiber links may be adopted in the future. Unfortunately, the overhead fiber links are dramatically affected by ambient temperature variation and mechanical stress. Consequently, they are much noisier than fiber spools and buried fiber [15-22]. Take SKA South Africa site for an example, it is constructed in the desert region. From the meteorological parameters captured by local weather station, rapid temperature change caused by thunderstorm, gale, sunrise and sunset is recorded frequently in the past. According to statistical analysis on the temperature data in the period from January 1, 2005 to March 31, 2011, the daily temperature fluctuation can reach 40°C [23]. Therefore, a robust frequency dissemination system immune to dramatic temperature fluctuation is required.

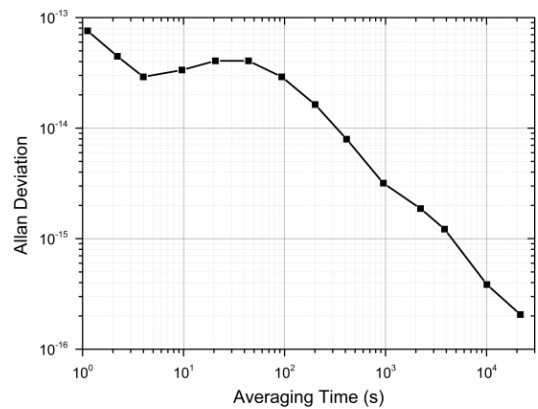

Fig. 1. The frequency dissemination stability result through 64 km overhead fiber at SKA South African site.

After performance test in the lab, the harmonic system was shipped to SKA South Africa site to perform an outfield test on several overhead fiber links from 19th to 27th September 2015. The frequency dissemination stability result via the 64 km overhead fiber is shown in Fig.1. It can be seen that there is a

This work was supported by the Program of International S&T Cooperation (No. 2016YFE0100200) and National Key Project of Research and Development (No. 2016YFA0302102).

Xi Zhu, Bo Wang, Yichen Guo, and Lijun Wang are with State Key Laboratory of Precision Measurement Technology and Instruments, Department of Precision Instrument. Tsinghua University, Beijing, 100084, China. (e-mail: bo.wang@tsinghua.edu.cn).

Xi Zhu, Yibo Yuan, and Lijun Wang are with Department of Physics, Tsinghua University, Beijing, 100084, China

Romeo Gamatham and Bruce Wallace are with SKA South Africa, Blend on Baker, 17 Baker Street, Rosebank 2196, Johannesburg, South Africa.

Keith Grainge is with Jodrell Bank Centre for Astrophysics, Alan Turing Building, School of Physics & Astronomy, The University of Manchester, Oxford Road, Manchester M13 9PL, UK.

bump on the Allan deviation plot of dissemination stability at the averaging time between 10 s and 100 s, which had never shown up in previous lab tests using fiber spools. While through theoretical analysis and experimental verification as described later, we find that the harmonic system is not perfectly immune to the dramatic temperature change of fiber links so that the phase fluctuation cannot be completely compensated.

In this paper, we analyze the reason why the previous harmonic scheme is not completely immune to temperature fluctuation and attribute this imperfection to limited isolation and nonlinear performance of the RF components in the system. According to the analyzation, we design a non-harmonic system and test its performance in an environment with temperature fluctuating 40 per day, which is similar to the ambient temperature fluctuation of SKA South Africa site. The result of this test shows a frequency dissemination stability of $4.0 \times 10^{-14}$/s and $3.0 \times 10^{-16}/10^4$s. Furthermore, we carry out another test under the situation of rapid temperature change – 20 per hour – and obtain a frequency dissemination stability of $3.0 \times 10^{-14}$/s and $3.0 \times 10^{-16}/10^4$s. Based on these test results, we conclude that the non-harmonic frequency dissemination system is robust with both huge and rapid temperature fluctuation, and hence is capable of operating in severe environment of large-scale scientific and engineering projects.

## II. SYSTEM AND THERMAL TESTING RESULTS

In 2015, we proposed a harmonic reference frequency synchronization scheme, as Fig. 2 shows. One transmitting site (TX) consists several extensible disseminating channels, so that it has the ability to simultaneously collaborate with several receiving sites (RXs). Here we just take one channel for example to explain the concept. At TX, the 100 MHz frequency signal from an H-maser acts as the reference signal. A phase-locked dielectric resonant oscillator (PDRO) with frequency of 2 GHz is phase locked to it and can be expressed as

$$V_0 = \cos(\omega_0 t + \phi_0). \tag{1}$$

It is used to modulate the amplitude of a 1547.72 nm diode laser (Laser A). Then the modulated laser light is divided into several equal light beams. After passing through a fiber coupler and an optical circulator, the light beam is disseminated from TX to corresponding RX via fiber link. As a performance test, a 50 km fiber spool is used. To simulate ambient temperature fluctuating, the fiber spool is placed in a temperature-controlled box. Their heating time can be adjusted through a digital PID control program developed with the LabVIEW interface. At RX, in order to actively compensate phase noise accumulated during fiber dissemination, another PDRO with frequency of 1 GHz is phase locked to a 100 MHz oven-controlled crystal oscillator (OCXO). The 1 GHz signal can be expressed as

$$V_1 = \cos(\omega_1 t + \phi_1) \tag{2}$$

and used to modulate the amplitude of a 1548.53 nm diode laser (Laser B). After passing through an optical circular, the modulated laser light is coupled into the same fiber link disseminating from RX to TX then back. At RX, the disseminated laser lights are separated from each other by a wavelength-division multiplexer (WDM). Two photodiodes (PDs) are used to recover the disseminated frequency signals, which can be expressed as

$$V_2 = \cos(\omega_0 t + \phi_0 + \phi_p); \tag{3}$$

$$V_3 = \cos(\omega_1 t + \phi_1 + \phi'_p). \tag{4}$$

Here, $\phi_p$ represents the phase fluctuation induced by 50 km fiber dissemination for the 2 GHz signal $V_0$ and $\phi'_p$ represents that of 50 km fiber round-trip dissemination for the 1 GHz signal $V_1$. The one-way accumulated phase fluctuation of the 2 GHz frequency signal is the same as the round-trip phase fluctuation of the 1 GHz frequency signal, namely $\phi_p = \phi'_p$. By frequency mixing down $V_2$ and $V_3$, we can get a 1 GHz signal:

$$V_4 = \cos[(\omega_0 - \omega_1)t + \phi_0 + \phi_p - \phi_1 - \phi'_p]. \tag{5}$$

Then $V_4$ is frequency mixed with $V_1$, and the down converted DC signal can be expressed as

$$V_5 = \cos[(\omega_0 - 2\omega_1)t + \phi_0 + \phi_p - 2\phi_1 - \phi'_p]. \tag{6}$$

As an error signal, $V_5$ is used to feedback-control the phase of the OCXO by a phase locked loop (PLL). With the relations of $\omega_0 = 2\omega_1$ and $\phi_p = \phi'_p$, $V_5$ also can be expressed as

$$V_5 = \cos(\phi_0 - 2\phi_1). \tag{7}$$

When the PLL is closed, the OCXO at RX is phase locked to the reference signal at TX, namely $\phi_0 = 2\phi_1$. Thus the reference signal is recovered at RX.

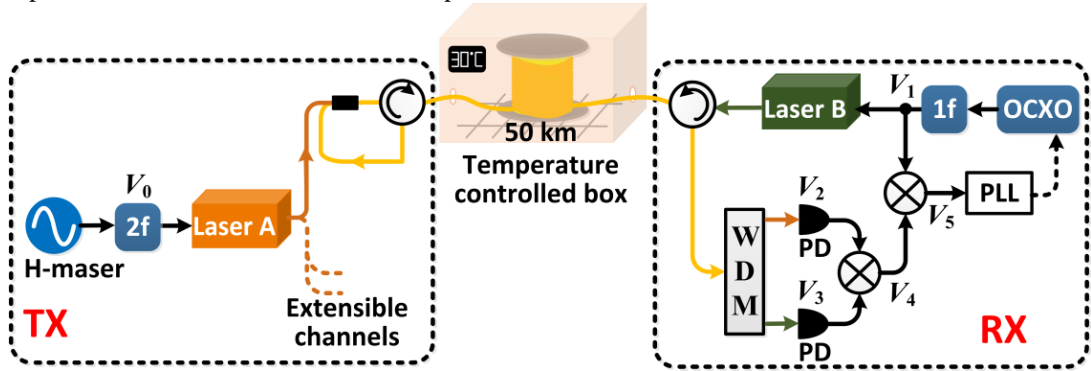

Fig. 2. Schematic diagram of the harmonic reference frequency synchronization scheme. OCXO: oven-controlled crystal oscillator, WDM: wavelength-division multiplexer, PD: photodiode, PLL: phase locked loop.

To simulate the ambient temperature fluctuation of SKA South Africa site, the temperature fluctuations of the temperature-controlled box are set to 10, 20, 30 and 40°C per day, respectively, as Fig. 3(a) shows. The corresponding relative frequency stability between the 100 MHz reference signal and the recovered 100 MHz signal is shown in Fig. 3(b). We can see from the results, when the ambient temperature of fiber link fluctuates, a bump shows up on the Allan deviation plot at the averaging time between 10 s and 1000 s. And the position of bump varies according to different temperature fluctuation.

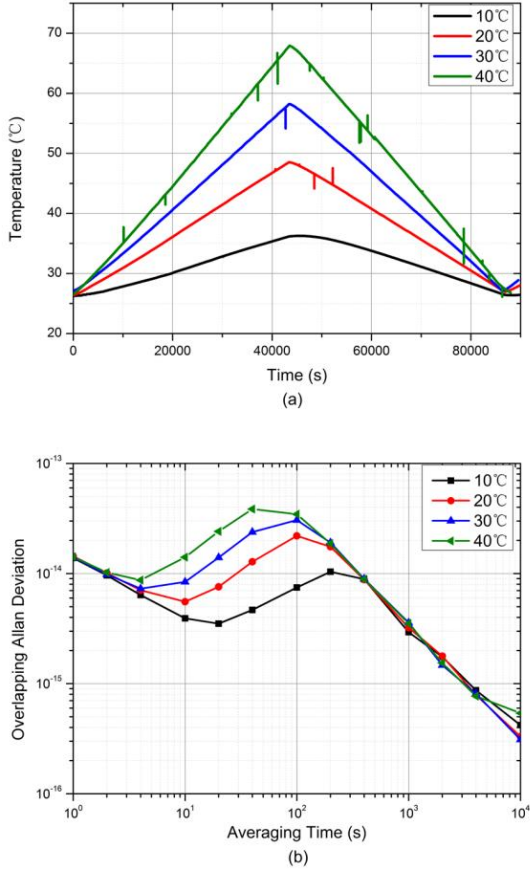

Fig. 3. (a) The ambient temperature of fiber link fluctuating 10, 20, 30 and 40°C per day. (b) Measured frequency dissemination stability with dramatic temperature fluctuation.

To explain this phenomenon, we further analysis the harmonic frequency synchronization scheme. In the previous discussion, the frequency mixing between $V_2$ and $V_3$ is considered to be an ideal process. That is to say, the output signal $V_4$ just contains the ideal signal (see (5)). Nevertheless, influenced by the limited isolation and nonlinear performance of frequency mixer in the system, components of the output signal $V_4$ are complex. Owing to the desired output signal with the frequency of 1 GHz, the leakage of $V_3$ with the same frequency cannot be filtered. It would limit the frequency dissemination stability. As to nonlinear performance, the frequency mixer we used (Marki T3-03) is featured of high IIP3 that can reduce the unwanted nonlinear effect, but the influence still exists. The frequency signals $V_2$ and $V_3$ in Taylor's series can be expressed as:

$$V_2 = \cos(\omega_0 t + \phi_0 + \phi_p) + a_1 \cos(2\omega_0 t + 2\phi_0 + 2\phi_p) \\ + a_2 \cos(3\omega_0 t + 3\phi_0 + 3\phi_p) + \cdots \quad (8)$$

and

$$V_3 = \cos(\omega_1 t + \phi_1 + \phi'_p) + b_1 \cos(2\omega_1 t + 2\phi_1 + 2\phi'_p) \\ + b_2 \cos(3\omega_1 t + 3\phi_1 + 3\phi'_p) + \cdots \quad (9)$$

where $a$ and $b$ are coefficients. The components of these two signals are frequency mixed with each other. Through a 1 GHz band pass filter, the output signal $V_4$ can be expressed as:

$$V_4 = \cos[(\omega_0 - \omega_1)t + \phi_0 + \phi_p - \phi_1 - \phi'_p] \\ + \xi \cos(\omega_1 t + \phi_1 + \phi'_p) \\ + \zeta \cos[(3\omega_1 - \omega_0)t + 3\phi_1 + 3\phi'_p - \phi_0 - \phi_p] \\ + \cdots \quad (10)$$

where $\xi$ and $\zeta$ are small quantities. The first component is the ideal signal, the second is the leakage of $V_3$ and the third is produced by 3rd harmonic of $V_3$ frequency mixing with $V_2$. Since other ultra-harmonics and products are much smaller, we just take these two influence factors into consideration. Considering $\phi_p = \phi'_p$, $V_5$ can be expressed as

$$V_5 = \cos(\phi_0 - 2\phi_1) \\ + \xi \cos(\phi_p) \\ + \zeta \cos(2\phi_1 + 2\phi_p - \phi_0) \quad (11)$$

The continuous variation of $\phi_p$ will lead to periodic variation of the second and third items of $V_5$ in (11). Thus the phase item $\phi_1$, which is tunable by changing the control voltage of the OCXO, will also vary periodically when the PLL is closed, instead of being completely equal to $\frac{1}{2}\phi_0$ as expected. Consequently, the Allan deviation plot of the frequency dissemination stability in Fig. 3 has a bump on it.

The fiber we used in lab is jacketed fiber. Considering the mechanical tension coefficients of fiber length and refractive index, the effective temperature coefficient of phase time delay is 76 $\frac{ps}{km \cdot °C}$ [24]. For the signal frequency of 2 GHz and the fiber length of 50 km, the temperature coefficient of phase delay should be

$$\frac{d\phi_p}{dT} = 15.2\pi \ rad/°C. \quad (12)$$

Since experiments show that the power of 3rd harmonic product component is ten times smaller than that of the 1 GHz leakage component, we will mainly consider the 1 GHz leakage in the analysis. The phase item $\phi_1$ will be disturbed at the same period with $\cos(\phi_p)$.

Taking temperature fluctuating 40°C/day as an example, the temperature changing rate is $\frac{dT}{dt}=\frac{1}{1080}\ °C/s$ (increasing or decreasing 40°C in 12 hours). According to (12), the phase delay changing rate can be calculated as:

$$\frac{d\phi_p}{dt}=\frac{d\phi_p}{dT}\cdot\frac{dT}{dt}=15.2\pi\cdot\frac{1}{1080}=0.014\pi\ rad/s \quad (13)$$

Correspondingly the period of $\cos(\phi_p)$ is $\frac{2\pi}{0.014\pi}=140\ s$, which is the same with the disturbance period of $\phi_1$. Consequently, the horizontal position of the bump on the Allan deviation plot should be at half of the period, which is 70 s. As to temperature fluctuating 10, 20 and 30 degrees, the horizontal positions should be 280 s, 140 s and 93 s, respectively. These calculated results agree well with the test results shown in Fig. 3(b).

### III. NON-HARMONIC SYSTEM AND RESULTS

The schematic diagram of non-harmonic precision reference frequency synchronization scheme is shown in Fig. 4. TX remains the same. Namely the expression of $V_0$ is the same as (1). After fiber dissemination from TX to RX, the expression of $V_2$ is also the same as in (3). At RX, the 1 GHz signal from the PDRO of previous scheme is replaced by two signals $1f_a$ and $1f_b$, with frequency of 1 GHz+130 Hz and 1 GHz-130 Hz respectively. Both of them are phase locked to the same OCXO. Then $V_1$ can be expressed as:

$$V_1=\cos\left[(\omega_1+130Hz)t+\phi_1\right], \quad (14)$$

and $V_3$ can be expressed as:

$$V_3=\cos\left[(\omega_1+130Hz)t+\phi_1+\phi'_p\right]. \quad (15)$$

By frequency mixing down $V_2$ and $V_3$, the down-conversion signal can be expressed as:

$$\begin{aligned}V_4=&\cos\left[(\omega_0-\omega_1-130Hz)t+\phi_0+\phi_p-\phi_1-\phi'_p\right]\\&+\xi\cos\left[(\omega_1+130Hz)t+\phi_1+\phi'_p\right]\\&+\zeta\cos\left[(3\omega_0-\omega_1+390Hz)t+3\phi_1+3\phi'_p-\phi_0-\phi_p\right]\end{aligned} \quad (16)$$

The second component is the leakage of $V_3$ and the last is produced by 3 rd. harmonic of $V_3$ mixing with $V_2$. The frequency signal $1f_b$ is used for frequency mixing, and it can be expressed as:

$$V_6=\cos\left[(\omega_1-130Hz)t+\phi_6\right], \quad (17)$$

where $\phi_6$ can be considered the same as $\phi_1$ with negligible difference. Then $V_6$ is mixed with $V_4$. An error signal

$$\begin{aligned}V_5=&\cos\left(\phi_0+\phi_p-2\phi_1-2\phi'_p\right)\\&+\xi\cos\left(260Hz\times t+\phi'_p\right)\\&+\zeta\cos\left(390Hz\times t+3\phi_1+3\phi'_p-\phi_0-\phi_p\right)\end{aligned} \quad (18)$$

is obtained. Through a low pass filter, the last two signal components can be almost filtered. In the non-harmonic scheme, the relation becomes $(1+1.3\times10^{-7})\phi_p=\phi'_p$. Then $V_5$ can be expressed as:

$$V_5=\cos\left(\phi_0-2\phi_1-1.3\times10^{-7}\phi_p\right). \quad (19)$$

When the PLL is closed, we have:

$$\phi_0=2\phi_1+1.3\times10^{-7}\phi_p. \quad (20)$$

A little part of $\phi_p$ is brought into the error signal. Taking temperature fluctuating 40°C as an example, the residual phase error is at the magnitude of $10^{-14}$ s, and it is negligible.

With the ambient temperature of fiber link fluctuating 40°C/day, which is similar to the temperature fluctuation of SKA South Africa site, we measure relative frequency stabilities between 100 MHz reference signal and recovered signal after a 50 km distance dissemination. The result is shown in Fig. 5(a). The bump on the Allan deviation plot of dissemination stability at the averaging time between 10 s and 1000 s almost diminished, and relative frequency stabilities of $4.0\times10^{-14}$/s and $3.0\times10^{-16}/10^4$ s are obtained.

Furthermore, to test the performance of this non-harmonic system under rapid temperature fluctuation, we conduct another experiment in which the ambient temperature of the fiber spool varies 20 °C/hour. There is also no bump on the Allan deviation plot and the relative frequency stability reaches $3.0\times10^{-14}$/s and $3.0\times10^{-16}/10^4$ s as shown in Fig. 5(b).

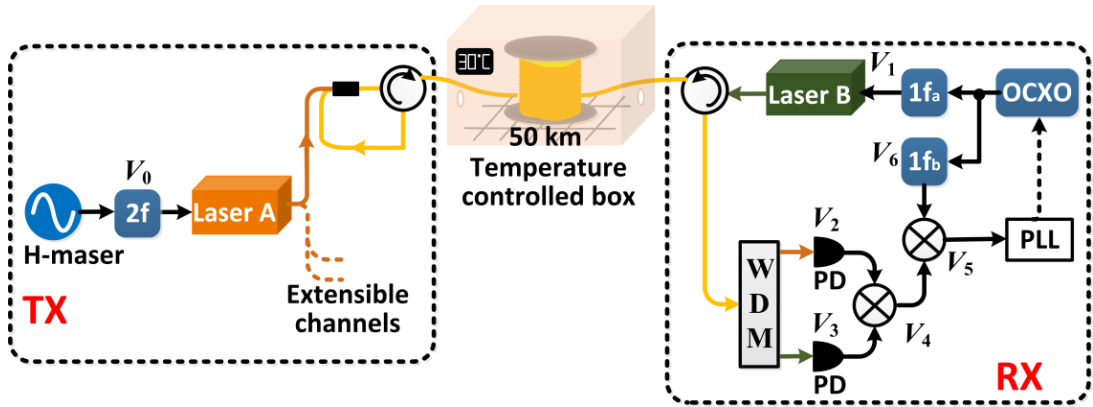

Fig. 4. The schematic diagram of non-harmonic precision reference frequency synchronization scheme. OCXO: oven-controlled crystal oscillator. WDM: wavelength-division multiplexer. PD: photodiode. $1f_a$: 1 GHz+130Hz signal phase locked to OCXO. $1f_b$: 1 GHz-130 Hz signal phase locked to OCXO.

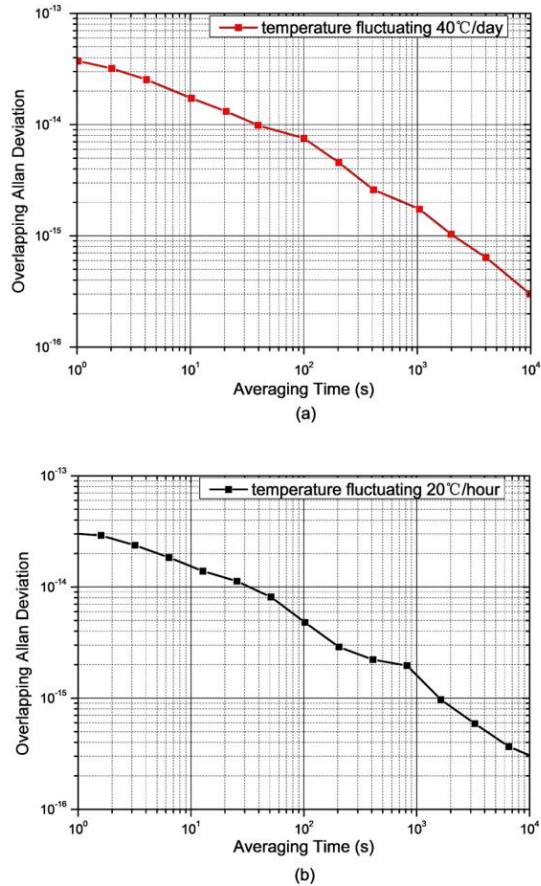

Fig. 5. Measured frequency dissemination stabilities of non-harmonic system for 50 km distance with temperature fluctuating of (a) 40°C/day, (b) 20°C/hour.

## IV. Conclusion

In summary, through outfield test in South Africa SKA site and thermal test in the lab, the harmonic frequency synchronization scheme is imperfect when using in environment with dramatic temperature fluctuation. The frequency synchronization stability is impacted by limited isolation and nonlinear performance of frequency mixer in the system. We design a non-harmonic synchronization scheme and prove it to be immune to dramatic temperature fluctuations, and hence is capable of operating in severe environment of large-scale scientific and engineering projects.

Acknowledgment

This work is being carried out for the SKA Signal and Data Transport (SaDT) consortium as part of the SKA project. The SKA project is an international effort to build the world's largest radio telescope, led by the SKA Organization with the support of 10 member countries. Fourteen institutions from eight countries are involved in the SaDT consortium, led by the University of Manchester.